\documentclass[sigconf]{acmart}

\AtBeginDocument{%
  \providecommand\BibTeX{{%
    \normalfont B\kern-0.5em{\scshape i\kern-0.25em b}\kern-0.8em\TeX}}}
\usepackage{color,xcolor}
\usepackage{comment}
\usepackage{utfsym}
\usepackage{algorithm}
\usepackage{subfigure}
\usepackage{algorithmic}
\usepackage{amsfonts}
\usepackage{soul}
\usepackage{amsmath}
\usepackage{multirow}
\usepackage{hyperref}
\hypersetup{
  colorlinks,
  citecolor=black,
  linkcolor=black,
  urlcolor=black}
\usepackage{caption}
\usepackage{colortbl}

\copyrightyear{2023}
\acmYear{2023}
\setcopyright{acmlicensed}\acmConference[IGSC '23]{THE 14th international Green and Sustainable Computing Conference}{October 28--29, 2023}{Toronto, ON, Canada}
\acmBooktitle{THE 14th international Green and Sustainable Computing Conference (IGSC '23), October 28--29, 2023, Toronto, ON, Canada}
\acmPrice{15.00}
\acmDOI{10.1145/3634769.3634801}
\acmISBN{979-8-4007-1669-0/23/10}

\begin{document}

\title{Hot-LEGO: Architect Microfluidic Cooling Equipped 3DICs with Pre-RTL Thermal Simulation}

\author{Runxi Wang}
\email{wangrunxi@sjtu.edu.cn}
\orcid{0009-0009-9615-9166}
\affiliation{%
  \institution{Shanghai Jiao Tong University}
  \city{Shanghai}
  \country{China}
}
\author{Jun-Han Han}
\email{jh2vs@virginia.edu}
\orcid{0009-0003-4317-7383}
\affiliation{%
  \institution{University of Virginia}
  \city{Charlottesville}
  \state{VA}
  \country{USA}
}
\author{Mircea Stan} 
\email{mircea@virginia.edu}
\orcid{0000-0003-0577-9976}
\affiliation{%
  \institution{University of Virginia}
  \city{Charlottesville}
  \state{VA}
  \country{USA}
}
\author{Xinfei Guo}
\authornote{Corresponding author.}
\email{xinfei.guo@sjtu.edu.cn}
\orcid{0000-0002-2374-3953}
\affiliation{%
  \institution{Shanghai Jiao Tong University}
  \city{Shanghai}
  \country{China}
}
\renewcommand{\shortauthors}{Runxi Wang et al.} 
\begin{abstract}
Microfluidic cooling has been recognized as one of the most promising solutions to achieve efficient thermal management for three-dimensional integrated circuits (3DICs). It enables more opportunities to architect 3DICs with different die configurations. It becomes increasingly important to perform thermal analysis in the early design phases to validate the architectural design decisions. This is even more critical for microfluidic cooling equipped 3DICs as the embedded cooling structures greatly influence the performance, power, and reliability of the stacked system. We exploited the existing architectural simulators and developed a Pre-register-transfer-level (Pre-RTL) thermal simulation methodology named Hot-LEGO that integrates these tools with their latest features such as support for microfluidic cooling and 3DIC stacking configurations. This methodology differs from existing ones by looking into the design granularity at a much finer level which enables the exploration of unique architecture combinations across the vertical stack. Though architectural-level simulators are not designed for signoff-calibre, it offers speed and agility which are imperative for early design space exploration. We claim that this ongoing work will speed up the co-design cycle of microfluidic cooling and offer a portable methodology for architects to perform exhaustive search for the optimal microarchitecture solutions in 3DICs.
\end{abstract}




\settopmatter{printacmref=false}
\setcopyright{none}
\renewcommand\footnotetextcopyrightpermission[1]{}
\pagestyle{plain}
\maketitle

\setlength\textfloatsep{0.4\baselineskip}
\setlength{\parskip}{0cm}
\section{Background \& Motivation}

\begin{figure}[t]
    \centering
    \includegraphics[width=\linewidth]{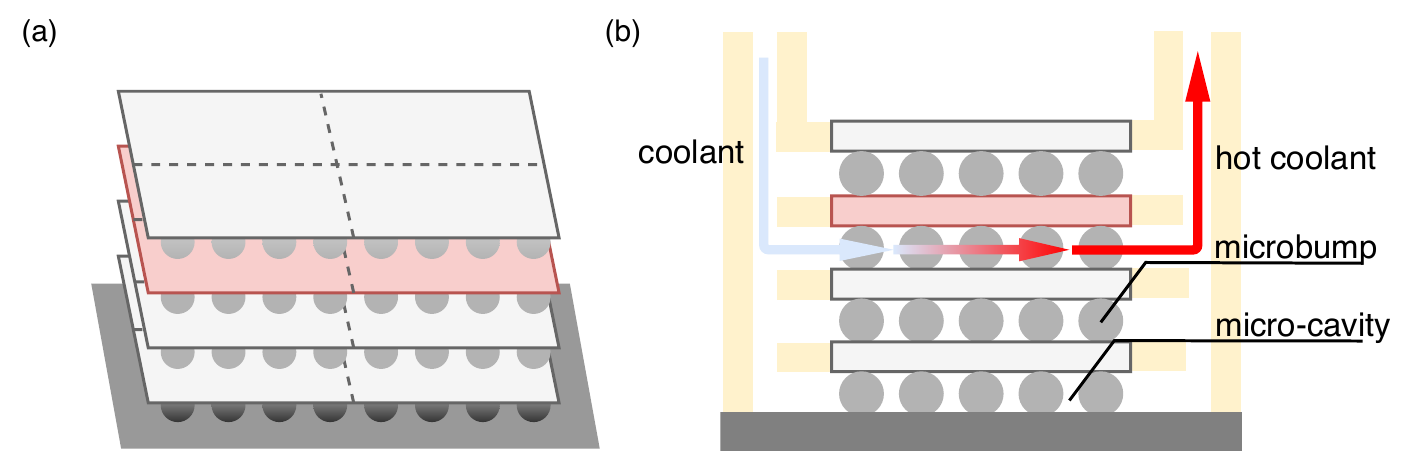}
    \caption{(a) Normal 3DIC stack; (b) Basic mechanisms of microfluidic cooling for 3DIC \cite{microfluidic}}
    \label{microfluidic}
\end{figure}

Thermal issue is one of the major road blockers that challenge the three-dimensional integrated circuits (3DIC) design due to increased power density and thermal resistance of the dielectric layers between the active devices \cite{leduc2007challenges,cao2019survey}. Typically, thermal analysis is done at the package or board level, resulting in irreversible design decisions if severe thermal issues are discovered later. This becomes more prominent for 3DIC, where multiple dies stack to push the limits of high-performance computing. Thermal analysis thus needs to be ``shifted left'' and performed early before the design process \cite{tavakkoli2016analysis,wan2014co}. Traditional multiphysics-based thermal simulation tools incur long simulation time and require many design details that are usually missing in the early design phase. And current ASIC design flows have little support for 3DIC thermal analysis. 
This necessity is exaggerated if novel cooling structures such as microchannels are included. Microfluidic cooling is a novel cooling method advantageous for 3D chip systems and has been validated in previous works such as \cite{microfluidic,wang20183d,9320506}. 
It enables the coolant to flow through the 3D stacks to cool down the chip as shown in Fig. \ref{microfluidic}.
This makes them more effective than traditional surface-based cooling techniques. So it offers more possibilities for designers and architects in terms of stacking configurations if codesign is feasible. In addition, the interlayer coupling with cooling structures, and a higher degree of connectivity among components, strengthen the interdependence between physical design parameters, architectural parameters, and a multitude of metrics of interest such as performance, hotspot distribution, and power \cite{wang20183d}. \par

The key challenges of thermal analysis in an early design phase come from two aspects. Firstly, how to build abstract models that capture the thermal details without losing accuracy. Secondly, how to perform thermal analysis on a ``pre-mature'' design efficiently so that design space exploration is feasible. The first challenge was addressed by recently proposed thermal simulation models that captured the physical parameters and material properties \cite{wan2014co,han2021thermal,wang20183d}. We explored various options and found that HotSpot \cite{skadron2002hotspot} offers a great balance of accuracy vs. efficacy. It was proposed in 2023 and has evolved to the 7th generation at this point \cite{stan2022}. In the latest version, it supports simulating microfluidic cooling after extending the thermal model to enable modeling heat transfer due to convection \cite{han2021thermal}. It is a widely-used architectural-level thermal simulator that requires very little information inputs, making it suitable for the pre-RTL design space exploration. The focus and contribution of our work is to address the second challenge by leveraging the power of HotSpot. Since thermal is highly relevant to power, it is required to develop a methodology that chains all these tools and explores an efficient way to expose the most optimal design option(s). In our proposed methodology Hot-LEGO, we aim to answer two questions, the first question is how will microfluidic cooling impact the stack configurations of a 3DIC. Here we are looking into ambitious stacking such as layered computing dies on top of each other, or sandwiched stacking where the computing dies are in between the memory dies. The second question is how to architect a single die so that its performance is maximized under the new cooling structures. This requires us to look into microarchitecture details and perform thermal simulation at a fine granularity level, such as at the cache level or ALU level. 
\begin{figure}[t]
    \centering
    \includegraphics[width=\linewidth]{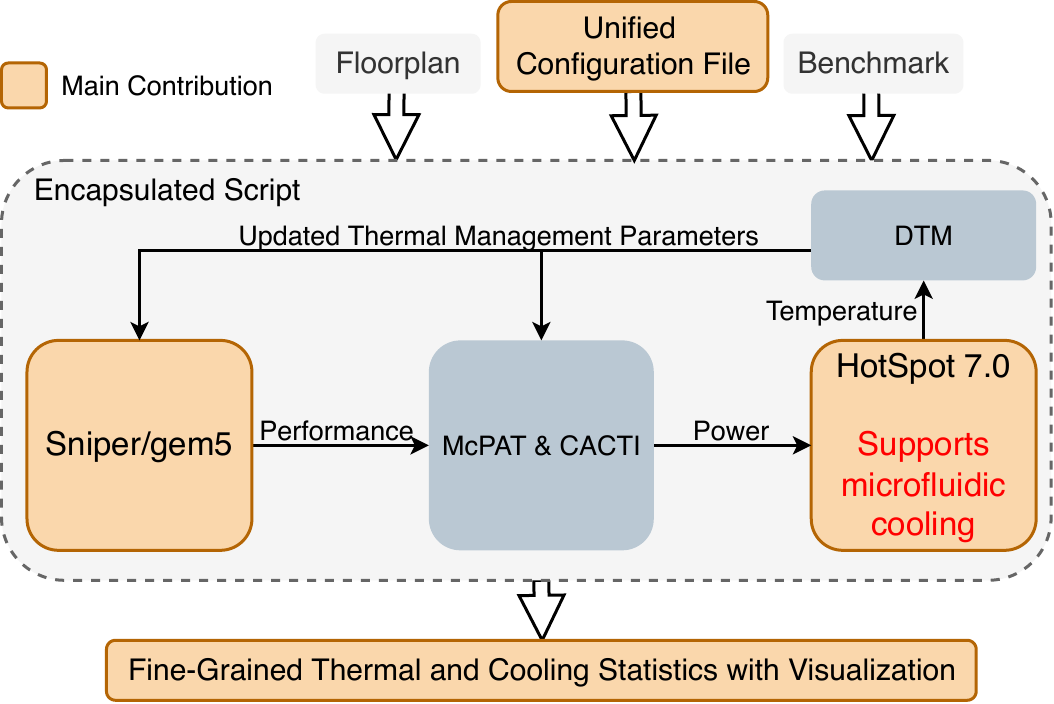}
    \vspace{-0.2cm}
    \caption{Hot-LEGO simulation framework and its main features.}
    \label{flow}
\end{figure}

\begin{figure*}[tbp!]
    \centering
    \vspace{-0.4cm}
    \includegraphics[width=0.495\linewidth]{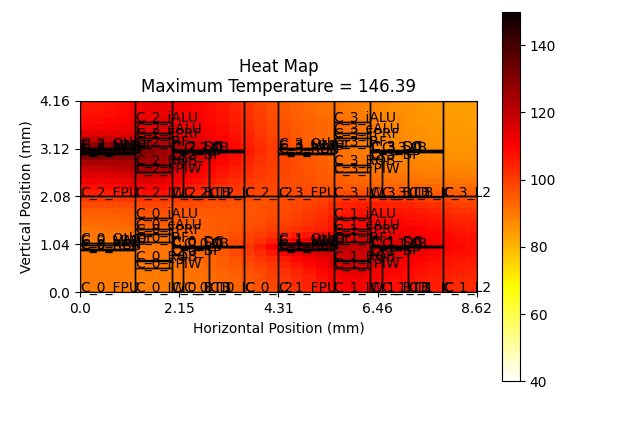}
    \includegraphics[width=0.495\linewidth]{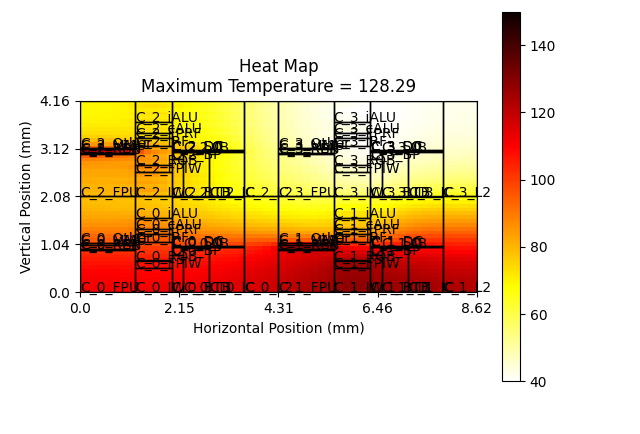}
    \vspace{-0.8cm}
    \caption{Fine-grained heat map of a core layer with normal configuration (left) and microfluidic cooling equipped (right)}
    \label{test}
\end{figure*}
\begin{figure*}[tbp!]
    \centering
    \includegraphics[width=0.495\linewidth]{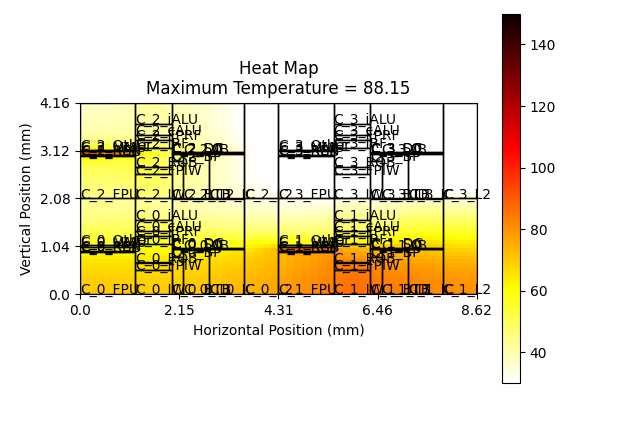}
    \includegraphics[width=0.495\linewidth]{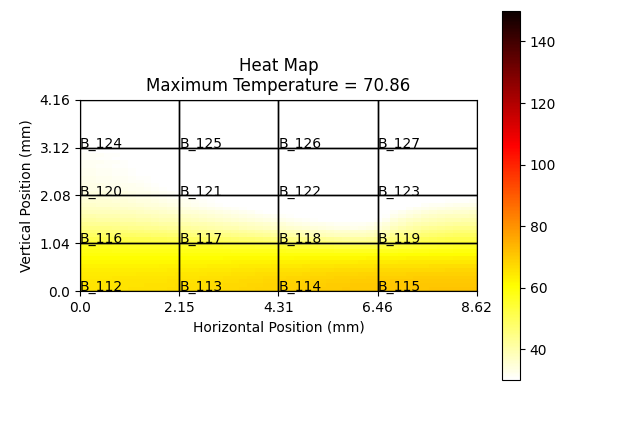}
    \vspace{-0.8cm}
    \caption{Heat map of a core layer (left) and a memory layer (right) in a 3DIC equipped with two microfluidic cooling layers}
    \label{test2}
\end{figure*}
\section{Hot-LEGO Simulation Framework}


Existing computer architecture simulators evaluate a design mainly on its power, performance, and area (PPA). 
For example, Gem5 \cite{binkert2011gem5,gem5link} is a performance simulator where one can customize the architectural design and check the architectural behaviors and latency with specified workloads. Sniper \cite{carlson2011sniper,snipertool} is another performance simulator that has more support on multi-core systems. For power simulators, CACTI \cite{shivakumar2001cacti,cactitool} and McPAT \cite{li2009mcpat,macpat} can be used to generate power traces. And for general thermal simulations, HotSpot \cite{skadron2002hotspot} provides fast and accurate thermal models. These simulators can be integrated together to provide an agile methodology for designers to explore design ideas in the Pre-RTL design phase. One example is CoMeT \cite{siddhu2022comet}, which targets thermal management simulations for various core-memory stacking configurations for 2D, 2.5D, and 3DICs. It has been used as an evaluation tool for several latest dynamic thermal management (DTM) research, such as \cite{siddhu2022corememdtm, pandey2022neuromap}. Though emerging simulation frameworks like CoMeT have raised attention to 3DIC, it still lacks simulation supports for advanced cooling methods like microfluidic cooling based on the thermal simulator they used. In addition, its simulation is limited to the whole core level without looking into microarchitectural details. 
The Hot-LEGO simulation framework discussed in this work is built on top of CoMeT but with significant extensions, such as extra support on microfluidic cooling strategy and support for checking thermal behaviors at finer design granularity. In addition, Hot-LEGO will enable efficient design space search at a larger scale by overcoming the interface limitations among different tools and tool versions. A diagram of the simulation framework is shown in Fig. \ref{flow} with added features highlighted. The framework mainly consists of three simulation stages, performance, power, and thermal stages. In the thermal simulation stage, we integrate the latest version of HotSpot which added the modeling for microfluidic cooling \cite{han20222,han2021thermal,hotspot7}. And the performance simulation will be enhanced to support more instruction set architectures (ISAs) using Gem5. Additionally, the microfluidic cooling is parameterized (such as injection rate) to enable the package-architecture codesign. To simplify the customization process, configurations of the cooling strategy will be integrated with other thermal parameters and structural parameters as a unified file. For outputs, the methodology will be capable of producing performance, area, power, and temperature statistics at a finer granularity with visualization support. Furthermore, different microarchitecture configurations will be explored to work with the novel cooling strategy. \par


In the current development phase, we are able to run several benchmarks from SPLASH-2 \cite{woo1995splash} and PARSEC \cite{bienia2008parsec} to compare hardware statistics under different cooling conditions. We show an example with PARSEC-vips in Fig. \ref{test}. To demonstrate the sole effect of microfluidic cooling, we turn off the default DTM policy in the configuration so that only the cooling support is made available for removing the heat. In the left diagram of Fig. \ref{test}, it is a heat map of processor cores without microfluidic cooling configuration, while in the right side, it is a heat map with one microfluidic cooling layer equipped near the bottom of the stack. From the reduced maximum temperature and the changed temperature distribution, the effectiveness of microfluidic cooling is shown. Different stacking manners of the microfluidic cooling layers can also contribute to better heat dissipation of the whole stack. We additionally add one microfluidic cooling layer with the same configurations near the top of the stack and keep other layers unchanged. The results are shown in Fig. \ref{test2}. For the core layer located on the bottom (Fig. \ref{test2} left), the cooling effects on it are enhanced compared with the case where there is only one cooling layer (Fig. \ref{test} right). For the memory layer located on top of the stack (Fig. \ref{test2} right), as it is at a better position to dissipate heat and with the equipment of the cooling layer neighboring to it, it can be seen that the temperature of it is much lower compared with the core layer. The darkness on the heat map reflects the thermal profile of different sub-blocks. \par
Currently, the framework is still in the preliminary phase, better visualization with more thermal metrics like energy efficiency, power dissipation, and cooling effect will be added. And stronger analysis support taking account 
of benchmark behaviors, stacking positions, cooling effects, and dynamic thermal management strategies will be equipped for better configuration guidance for designers. Through this type of study, one can take early actions that concern the thermal aspects even before the actual design phase starts. The analysis fosters the connections between the designers, packaging and board teams, offering perspectives to architect 3DIC as a whole. It is worth to mention that the accuracy of the analysis relies on the inherent modeling framework of each tool, which can be calibrated with actual measured results. 

\section{Summary and Outlook}
As a first-order mission-critical parameter, thermal analysis needs to be designed instead of being fixed after the fact. In this paper, we laid out our plan for developing a pre-RTL thermal simulation methodology which targets microfluidic cooling equipped 3DICs. Through this methodology, early design space exploration is made feasible by considering the cooling structures. To achieve these goals, we need to overcome a few key challenges. Firstly, we need to make the tradeoff between accuracy and efficacy. Pre-RTL simulations are based on abstract models, thus they are not designed to perform signoff analysis. In the meantime, they should reflect the ``actual'' trend with major tradeoffs built in. More calibration can be added to correlate the simulated results with the measured data. Secondly, this methodology is built on top of existing tools. Thus it will be limited by features available in these tools. We will explore the possibilities of enhancing these tools to support more features. For example, stacking memory and core layers together broadens the way to achieve in-memory or near-memory computing architecture design. With the methodology proposed, we are looking into simulating computing-in-memory architectures in order to explore a better combination of memory and core layers. Lastly, on-chip DTM policies need to work together with the cooling structures to achieve better performance. Microfluidic cooling will work more flexibly under the control from DTM policy. With this methodology, we will look into the confluence of cooling structures and DTM policies to find out a cost-effective solution in an early design phase. The goal is to establish an integrated methodology that assists 3DIC designers and architects to make quick yet accurate early design decisions before entering the design phase.



\begin{acks}
This work was partially supported by National Science Foundation of China under Grant No. 62201340. Authors would like to thank Gem5, Sniper, CACTI, McPAT and HotSpot developers and active contributors for providing the infrastructures to enable this research. 
\end{acks}

\newpage
\bibliographystyle{ACM-Reference-Format}
\balance
\bibliography{reference}


\begin{thebibliography}{25}


\ifx \showCODEN    \undefined \def \showCODEN     #1{\unskip}     \fi
\ifx \showDOI      \undefined \def \showDOI       #1{#1}\fi
\ifx \showISBNx    \undefined \def \showISBNx     #1{\unskip}     \fi
\ifx \showISBNxiii \undefined \def \showISBNxiii  #1{\unskip}     \fi
\ifx \showISSN     \undefined \def \showISSN      #1{\unskip}     \fi
\ifx \showLCCN     \undefined \def \showLCCN      #1{\unskip}     \fi
\ifx \shownote     \undefined \def \shownote      #1{#1}          \fi
\ifx \showarticletitle \undefined \def \showarticletitle #1{#1}   \fi
\ifx \showURL      \undefined \def \showURL       {\relax}        \fi
\providecommand\bibfield[2]{#2}
\providecommand\bibinfo[2]{#2}
\providecommand\natexlab[1]{#1}
\providecommand\showeprint[2][]{arXiv:#2}

\bibitem[mac(2013)]%
        {macpat}
 \bibinfo{year}{2013}\natexlab{}.
\newblock \bibinfo{booktitle}{\emph{McPAT}}.
\newblock
\urldef\tempurl%
\url{https://code.google.com/archive/p/mcpat/}
\showURL{%
\tempurl}


\bibitem[cac(2017)]%
        {cactitool}
 \bibinfo{year}{2017}\natexlab{}.
\newblock \bibinfo{booktitle}{\emph{CACTI}}.
\newblock
\urldef\tempurl%
\url{https://github.com/HewlettPackard/cacti}
\showURL{%
\tempurl}


\bibitem[hot(2022)]%
        {hotspot7}
 \bibinfo{year}{2022}\natexlab{}.
\newblock \bibinfo{booktitle}{\emph{HotSpot 7.0}}.
\newblock
\urldef\tempurl%
\url{https://github.com/uvahotspot/HotSpot}
\showURL{%
\tempurl}


\bibitem[gem(2023)]%
        {gem5link}
 \bibinfo{year}{2023}\natexlab{}.
\newblock \bibinfo{booktitle}{\emph{Gem5}}.
\newblock
\urldef\tempurl%
\url{https://www.gem5.org/}
\showURL{%
\tempurl}


\bibitem[sni(2023)]%
        {snipertool}
 \bibinfo{year}{2023}\natexlab{}.
\newblock \bibinfo{booktitle}{\emph{The Sniper Multi-Core Simulator}}.
\newblock
\urldef\tempurl%
\url{https://snipersim.org/w/The_Sniper_Multi-Core_Simulator}
\showURL{%
\tempurl}


\bibitem[Bienia et~al\mbox{.}(2008)]%
        {bienia2008parsec}
\bibfield{author}{\bibinfo{person}{Christian Bienia}, \bibinfo{person}{Sanjeev Kumar}, \bibinfo{person}{Jaswinder~Pal Singh}, {and} \bibinfo{person}{Kai Li}.} \bibinfo{year}{2008}\natexlab{}.
\newblock \showarticletitle{The PARSEC benchmark suite: Characterization and architectural implications}. In \bibinfo{booktitle}{\emph{Proceedings of the 17th international conference on Parallel architectures and compilation techniques}}. \bibinfo{pages}{72--81}.
\newblock


\bibitem[Binkert et~al\mbox{.}(2011)]%
        {binkert2011gem5}
\bibfield{author}{\bibinfo{person}{Nathan Binkert}, \bibinfo{person}{Bradford Beckmann}, \bibinfo{person}{Gabriel Black}, \bibinfo{person}{Steven~K Reinhardt}, \bibinfo{person}{Ali Saidi}, \bibinfo{person}{Arkaprava Basu}, \bibinfo{person}{Joel Hestness}, \bibinfo{person}{Derek~R Hower}, \bibinfo{person}{Tushar Krishna}, \bibinfo{person}{Somayeh Sardashti}, {et~al\mbox{.}}} \bibinfo{year}{2011}\natexlab{}.
\newblock \showarticletitle{The gem5 simulator}.
\newblock \bibinfo{journal}{\emph{ACM SIGARCH computer architecture news}} \bibinfo{volume}{39}, \bibinfo{number}{2} (\bibinfo{year}{2011}), \bibinfo{pages}{1--7}.
\newblock


\bibitem[Cao et~al\mbox{.}(2019)]%
        {cao2019survey}
\bibfield{author}{\bibinfo{person}{Kun Cao}, \bibinfo{person}{Junlong Zhou}, \bibinfo{person}{Tongquan Wei}, \bibinfo{person}{Mingsong Chen}, \bibinfo{person}{Shiyan Hu}, {and} \bibinfo{person}{Keqin Li}.} \bibinfo{year}{2019}\natexlab{}.
\newblock \showarticletitle{A survey of optimization techniques for thermal-aware 3D processors}.
\newblock \bibinfo{journal}{\emph{Journal of Systems Architecture}}  \bibinfo{volume}{97} (\bibinfo{year}{2019}), \bibinfo{pages}{397--415}.
\newblock


\bibitem[Carlson et~al\mbox{.}(2011)]%
        {carlson2011sniper}
\bibfield{author}{\bibinfo{person}{Trevor~E Carlson}, \bibinfo{person}{Wim Heirman}, {and} \bibinfo{person}{Lieven Eeckhout}.} \bibinfo{year}{2011}\natexlab{}.
\newblock \showarticletitle{Sniper: Exploring the level of abstraction for scalable and accurate parallel multi-core simulation}. In \bibinfo{booktitle}{\emph{Proceedings of 2011 International Conference for High Performance Computing, Networking, Storage and Analysis}}. \bibinfo{pages}{1--12}.
\newblock


\bibitem[Han et~al\mbox{.}(2022)]%
        {han20222}
\bibfield{author}{\bibinfo{person}{Jun-Han Han}, \bibinfo{person}{Xinfei Guo}, \bibinfo{person}{Kevin Skadron}, {and} \bibinfo{person}{Mircea~R Stan}.} \bibinfo{year}{2022}\natexlab{}.
\newblock \showarticletitle{From 2.5 D to 3D Chiplet Systems: Investigation of Thermal Implications with HotSpot 7.0}. In \bibinfo{booktitle}{\emph{2022 21st IEEE Intersociety Conference on Thermal and Thermomechanical Phenomena in Electronic Systems (iTherm)}}. IEEE, \bibinfo{pages}{1--6}.
\newblock


\bibitem[Han et~al\mbox{.}(2019a)]%
        {microfluidic}
\bibfield{author}{\bibinfo{person}{Jun-Han Han}, \bibinfo{person}{Karina Torres-Castro}, \bibinfo{person}{Robert~E. West}, \bibinfo{person}{Walter Varhue}, \bibinfo{person}{Nathan Swami}, {and} \bibinfo{person}{Mircea Stan}.} \bibinfo{year}{2019}\natexlab{a}.
\newblock \showarticletitle{Microfluidic Cooling for 3D-IC with 3D Printing Package}. In \bibinfo{booktitle}{\emph{2019 IEEE SOI-3D-Subthreshold Microelectronics Technology Unified Conference (S3S)}}. \bibinfo{pages}{1--3}.
\newblock
\urldef\tempurl%
\url{https://doi.org/10.1109/S3S46989.2019.9320506}
\showDOI{\tempurl}


\bibitem[Han et~al\mbox{.}(2019b)]%
        {9320506}
\bibfield{author}{\bibinfo{person}{Jun-Han Han}, \bibinfo{person}{Karina Torres-Castro}, \bibinfo{person}{Robert~E. West}, \bibinfo{person}{Walter Varhue}, \bibinfo{person}{Nathan Swami}, {and} \bibinfo{person}{Mircea Stan}.} \bibinfo{year}{2019}\natexlab{b}.
\newblock \showarticletitle{Microfluidic Cooling for 3D-IC with 3D Printing Package}. In \bibinfo{booktitle}{\emph{2019 IEEE SOI-3D-Subthreshold Microelectronics Technology Unified Conference (S3S)}}. \bibinfo{pages}{1--3}.
\newblock
\urldef\tempurl%
\url{https://doi.org/10.1109/S3S46989.2019.9320506}
\showDOI{\tempurl}


\bibitem[Han et~al\mbox{.}(2021)]%
        {han2021thermal}
\bibfield{author}{\bibinfo{person}{Jun-Han Han}, \bibinfo{person}{Robert~E West}, \bibinfo{person}{Kevin Skadron}, {and} \bibinfo{person}{Mircea~R Stan}.} \bibinfo{year}{2021}\natexlab{}.
\newblock \showarticletitle{Thermal simulation of processing-in-memory devices using HotSpot 7.0}. In \bibinfo{booktitle}{\emph{2021 27th International Workshop on Thermal Investigations of ICs and Systems (THERMINIC)}}. IEEE, \bibinfo{pages}{1--5}.
\newblock


\bibitem[Leduc et~al\mbox{.}(2007)]%
        {leduc2007challenges}
\bibfield{author}{\bibinfo{person}{Patrick Leduc}, \bibinfo{person}{Francois de Crecy}, \bibinfo{person}{Murielle Fayolle}, \bibinfo{person}{Barbara Charlet}, \bibinfo{person}{Thierry Enot}, \bibinfo{person}{Marc Zussy}, \bibinfo{person}{Bob Jones}, \bibinfo{person}{Jean-Charles Barbe}, \bibinfo{person}{Nelly Kernevez}, \bibinfo{person}{Nicolas Sillon}, {et~al\mbox{.}}} \bibinfo{year}{2007}\natexlab{}.
\newblock \showarticletitle{Challenges for 3D IC integration: bonding quality and thermal management}. In \bibinfo{booktitle}{\emph{2007 IEEE International Interconnect Technology Conferencee}}. IEEE, \bibinfo{pages}{210--212}.
\newblock


\bibitem[Li et~al\mbox{.}(2009)]%
        {li2009mcpat}
\bibfield{author}{\bibinfo{person}{Sheng Li}, \bibinfo{person}{Jung~Ho Ahn}, \bibinfo{person}{Richard~D Strong}, \bibinfo{person}{Jay~B Brockman}, \bibinfo{person}{Dean~M Tullsen}, {and} \bibinfo{person}{Norman~P Jouppi}.} \bibinfo{year}{2009}\natexlab{}.
\newblock \showarticletitle{McPAT: An integrated power, area, and timing modeling framework for multicore and manycore architectures}. In \bibinfo{booktitle}{\emph{Proceedings of the 42nd annual ieee/acm international symposium on microarchitecture}}. \bibinfo{pages}{469--480}.
\newblock


\bibitem[Pandey and Panda(2022)]%
        {pandey2022neuromap}
\bibfield{author}{\bibinfo{person}{Shailja Pandey} {and} \bibinfo{person}{Preeti~Ranjan Panda}.} \bibinfo{year}{2022}\natexlab{}.
\newblock \showarticletitle{NeuroMap: Efficient Task Mapping of Deep Neural Networks for Dynamic Thermal Management in High-Bandwidth Memory}.
\newblock \bibinfo{journal}{\emph{IEEE Transactions on Computer-Aided Design of Integrated Circuits and Systems}} \bibinfo{volume}{41}, \bibinfo{number}{11} (\bibinfo{year}{2022}), \bibinfo{pages}{3602--3613}.
\newblock


\bibitem[Shivakumar and Jouppi(2001)]%
        {shivakumar2001cacti}
\bibfield{author}{\bibinfo{person}{Premkishore Shivakumar} {and} \bibinfo{person}{Norman~P Jouppi}.} \bibinfo{year}{2001}\natexlab{}.
\newblock \showarticletitle{Cacti 3.0: An integrated cache timing, power, and area model}.
\newblock  (\bibinfo{year}{2001}).
\newblock


\bibitem[Siddhu et~al\mbox{.}(2022a)]%
        {siddhu2022corememdtm}
\bibfield{author}{\bibinfo{person}{Lokesh Siddhu}, \bibinfo{person}{Rajesh Kedia}, {and} \bibinfo{person}{Preeti~Ranjan Panda}.} \bibinfo{year}{2022}\natexlab{a}.
\newblock \showarticletitle{CoreMemDTM: Integrated processor core and 3D memory dynamic thermal management for improved performance}. In \bibinfo{booktitle}{\emph{2022 Design, Automation \& Test in Europe Conference \& Exhibition (DATE)}}. IEEE, \bibinfo{pages}{1377--1382}.
\newblock


\bibitem[Siddhu et~al\mbox{.}(2022b)]%
        {siddhu2022comet}
\bibfield{author}{\bibinfo{person}{Lokesh Siddhu}, \bibinfo{person}{Rajesh Kedia}, \bibinfo{person}{Shailja Pandey}, \bibinfo{person}{Martin Rapp}, \bibinfo{person}{Anuj Pathania}, \bibinfo{person}{J{\"o}rg Henkel}, {and} \bibinfo{person}{Preeti~Ranjan Panda}.} \bibinfo{year}{2022}\natexlab{b}.
\newblock \showarticletitle{CoMeT: An integrated interval thermal simulation toolchain for 2D, 2.5 D, and 3D processor-memory systems}.
\newblock \bibinfo{journal}{\emph{ACM Transactions on Architecture and Code Optimization (TACO)}} \bibinfo{volume}{19}, \bibinfo{number}{3} (\bibinfo{year}{2022}), \bibinfo{pages}{1--25}.
\newblock


\bibitem[Skadron et~al\mbox{.}(2002)]%
        {skadron2002hotspot}
\bibfield{author}{\bibinfo{person}{Kevin Skadron}, \bibinfo{person}{Mircea Stan}, \bibinfo{person}{Marco Barcella}, \bibinfo{person}{Amar Dwarka}, \bibinfo{person}{Wei Huang}, \bibinfo{person}{Yingmin Li}, \bibinfo{person}{Yong Ma}, \bibinfo{person}{Amit Naidu}, \bibinfo{person}{Dharmesh Parikh}, \bibinfo{person}{Paolo Re}, {et~al\mbox{.}}} \bibinfo{year}{2002}\natexlab{}.
\newblock \showarticletitle{HotSpot: Techniques for modeling thermal effects at the processor-architecture level}. THERMINIC.
\newblock


\bibitem[Stan et~al\mbox{.}(2022)]%
        {stan2022}
\bibfield{author}{\bibinfo{person}{Mircea~R Stan}, \bibinfo{person}{Kevin Skadron}, \bibinfo{person}{Xinfei Guo}, {and} \bibinfo{person}{Jun-Han Han}.} \bibinfo{year}{2022}\natexlab{}.
\newblock \showarticletitle{HotSpot Through the Ages}. In \bibinfo{booktitle}{\emph{HSSB: HotSpots Strike Back, co-located with ISCA}}. \bibinfo{pages}{1--2}.
\newblock
\urldef\tempurl%
\url{https://sites.tufts.edu/tcal/files/2022/06/ISCA22_HSSB_paper_3.pdf}
\showURL{%
\tempurl}


\bibitem[Tavakkoli et~al\mbox{.}(2016)]%
        {tavakkoli2016analysis}
\bibfield{author}{\bibinfo{person}{Fatemeh Tavakkoli}, \bibinfo{person}{Siavash Ebrahimi}, \bibinfo{person}{Shujuan Wang}, {and} \bibinfo{person}{Kambiz Vafai}.} \bibinfo{year}{2016}\natexlab{}.
\newblock \showarticletitle{Analysis of critical thermal issues in 3D integrated circuits}.
\newblock \bibinfo{journal}{\emph{International Journal of Heat and Mass Transfer}}  \bibinfo{volume}{97} (\bibinfo{year}{2016}), \bibinfo{pages}{337--352}.
\newblock


\bibitem[Wan et~al\mbox{.}(2014)]%
        {wan2014co}
\bibfield{author}{\bibinfo{person}{Zhimin Wan}, \bibinfo{person}{He Xiao}, \bibinfo{person}{Yogendra Joshi}, {and} \bibinfo{person}{Sudhakar Yalamanchili}.} \bibinfo{year}{2014}\natexlab{}.
\newblock \showarticletitle{Co-design of multicore architectures and microfluidic cooling for 3D stacked ICs}.
\newblock \bibinfo{journal}{\emph{Microelectronics Journal}} \bibinfo{volume}{45}, \bibinfo{number}{12} (\bibinfo{year}{2014}), \bibinfo{pages}{1814--1821}.
\newblock


\bibitem[Wang et~al\mbox{.}(2018)]%
        {wang20183d}
\bibfield{author}{\bibinfo{person}{Shaoxi Wang}, \bibinfo{person}{Yue Yin}, \bibinfo{person}{Chenxia Hu}, {and} \bibinfo{person}{Pouya Rezai}.} \bibinfo{year}{2018}\natexlab{}.
\newblock \showarticletitle{3D integrated circuit cooling with microfluidics}.
\newblock \bibinfo{journal}{\emph{Micromachines}} \bibinfo{volume}{9}, \bibinfo{number}{6} (\bibinfo{year}{2018}), \bibinfo{pages}{287}.
\newblock


\bibitem[Woo et~al\mbox{.}(1995)]%
        {woo1995splash}
\bibfield{author}{\bibinfo{person}{Steven~Cameron Woo}, \bibinfo{person}{Moriyoshi Ohara}, \bibinfo{person}{Evan Torrie}, \bibinfo{person}{Jaswinder~Pal Singh}, {and} \bibinfo{person}{Anoop Gupta}.} \bibinfo{year}{1995}\natexlab{}.
\newblock \showarticletitle{The SPLASH-2 programs: Characterization and methodological considerations}.
\newblock \bibinfo{journal}{\emph{ACM SIGARCH computer architecture news}} \bibinfo{volume}{23}, \bibinfo{number}{2} (\bibinfo{year}{1995}), \bibinfo{pages}{24--36}.
\newblock


\end{thebibliography}


\end{document}